\documentclass[fleqn,10pt]{wlscirep}

\newcommand{\mgion}{\ensuremath{^{25}\mathrm{Mg}^+}}

\title{Hybrid setup for stable magnetic fields enabling robust quantum control}

\author[1,*]{Frederick Hakelberg}
\author[1]{Philip Kiefer}
\author[1]{Matthias Wittemer}
\author[1]{Tobias Schaetz}
\author[1]{Ulrich Warring}
\affil[1]{Albert-Ludwigs-Universit\"at Freiburg, Physikalisches Institut, Hermann-Herder-Stra\ss e 3, 79104 Freiburg, Germany}

\affil[*]{frederick.hakelberg@physik.uni-freiburg.de}

\usepackage{textcomp}
\usepackage{gensymb}
\usepackage{xcolor}
{
}

\definecolor{myorange}{RGB}{255,127,14}
\definecolor{mygreen}{RGB}{44,160,44}
\definecolor{myred}{RGB}{214,39,40}
\definecolor{myviolet}{RGB}{148,103,189}
\definecolor{myblue}{RGB}{31,119,180}

\usepackage{setspace}
\begin{abstract}
	Well controlled and highly stable magnetic fields are desired for a wide range of applications in physical research, including quantum metrology, sensing, information processing, and simulation. 
	Here we introduce a low-cost hybrid assembly of rare-earth magnets and magnetic field coils to generate a field strength of $\simeq\,10.9\,$mT with a spatial variation of less than 10$^{-6}$ within a diameter of spherical volume of $150\,$\textmu m.
	We characterise its tuneability and stability performance using a single Mg$^{+}$ atom confined in a radio-frequency surface-electrode trap under ultra-high vacuum conditions. 
	The strength of the field can be tuned with a relative precision of $\leq 2\,\times\,10^{-5}$ and we find a passive temporal stability of our setup of better than $1.0\,\times\,10^{-4}$ over the course of one hour.
	Slow drifts on time scales of a few minutes are actively stabilised by adjusting electric currents in the magnetic field coils. 
	In this way, we observe coherence times of electronic superposition states of greater than six seconds using a first-order field insensitive (clock) transition.
	In a first application, we demonstrate sensing of magnetic fields with amplitudes of $\geq0.2\,$\textmu T oscillating at $\simeq 2\pi\,\times\,60\,$MHz.
	Our approach can be implemented in compact and robust applications with strict power and load requirements.
\end{abstract}
\begin{document}
	
	\flushbottom
	\maketitle
	\thispagestyle{empty}
	
	\section*{Introduction}
Quantum technologies\,\cite{Georgescu2012} are developed for a wide range of applications in the context of metrology\,\cite{Giovannetti2011}, sensing\,\cite{Degen2017}, information processing\,\cite{Ladd2010}, communication\,\cite{Gisin2007}, and simulation\,\cite{Buluta2009, Bloch2012, Schaetz2013, Georgescu2014}. 
While different experimental platforms are studied\,\cite{Xiang2013,Ladd2010,Buluta2009,Georgescu2014,Degen2017}, atomic systems, in particular, perform quantum gate operations with highest fidelities\,\cite{Ladd2010,Gaebler2016c,Ballance2016} and present clocks with exceptional precision\,\cite{Schmidt2005,Rosenband2008,Chou2010,Bloom2014,Huntemann2016,Zhang2016}.
Generally, advantageous performance of any quantum application in comparison to classical counterparts can be harnessed only when required control fields interplay with a high level of precision, while the system is well isolated from environmental disturbances. 
For example, static magnetic (quantisation) fields tune and stabilise electronic states of atoms to desired energy splittings which can be addressed by additional control fields for state manipulation.
Fidelities of coherent manipulations crucially depend on the performance of any such quantisation fields.
For some applications, specific combinations of atomic species and field strengths can be desired and enable the use of so-called first-order field insensitive (clock) transitions\,\cite{Bollinger1991a,Langer2005} that are less sensitive to field fluctuations than others.
Typically, fields are generated by field coils and to ensure stable operation conditions cooling and stable high-power current supplies are required. 
To further increase fidelities and complexity of quantum applications and/or to enable portable devices, robust and compact experimental setups with highly integrated components are required and being developed\,\,\cite{Ospelkaus2011,Cheng2016,Ruster2016c,Schwindt2016,Koller2017}. 
Under these circumstances the use of rare-earth magnets to create quantisation fields can be beneficial in contrast to field coils. 
In the last years, such permanent magnets became more popular for a variety of applications in atomic physics research\,\cite{Fernholz2008a,tan2012,Khromova2012,Lebedev2014,Kawai2017}, in particular, due to their high magnetisation and despite their limited tunability of field strengths.

In our manuscript, we introduce a hybrid approach, using an assembly of rare-earth magnets and pairs of field coils, to generate well-controlled quantisation fields with strengths of more than 10\,mT.
To benchmark the performance of our approach, we use a single trapped Mg$^+$ atom as a quantum sensor.
Further, we implement a protocol to probe stray magnetic fields with amplitudes of $\geq0.2\,$\textmu T oscillating at radio-frequencies enabled by the high stability of our magnetic field setup. 
	\section*{Experimental Setup}
\begin{figure}
		\centering
		\includegraphics[]{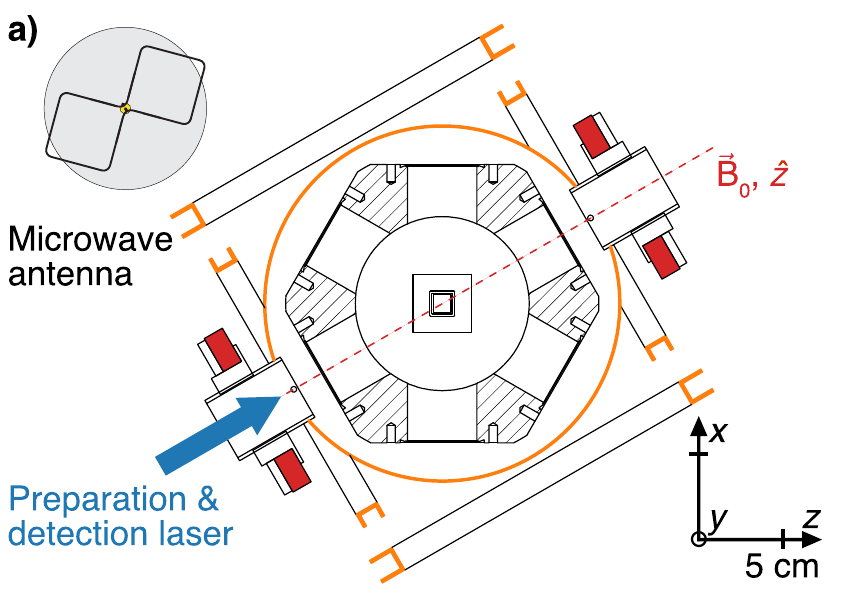}
		\includegraphics[]{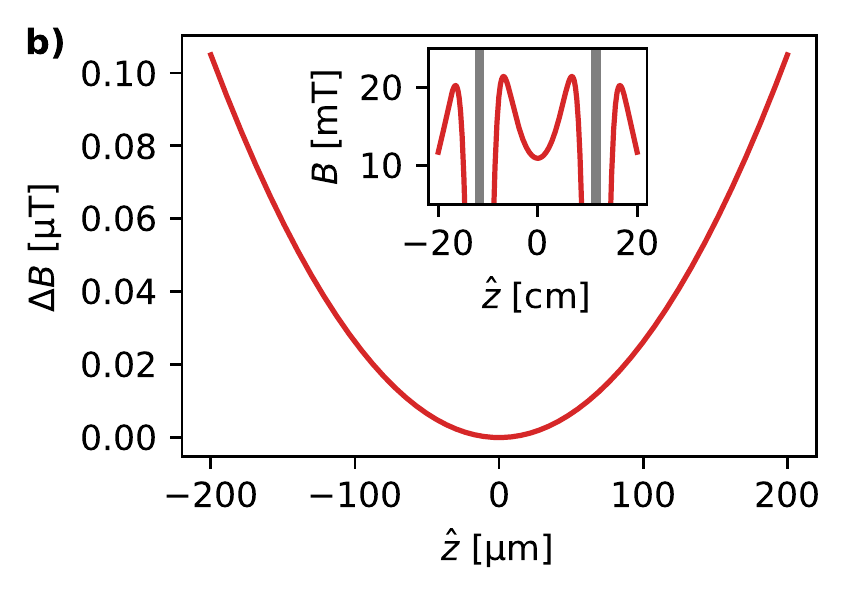}
		\caption{
		\textbf{Experimental setup and spatial properties of the solid-state magnet assembly.} 
		\textbf{(a)} Cross-sectional view of a ultra-high vacuum chamber housing a surface-electrode trap (indicated at the centre), used for spatial manipulation of single atoms. 
		Two sets of rare-earth, ring magnets generate a magnetic (quantisation) field $\mathbf{B}_0$ along their symmetry axis $\hat{z}$ (indicated by \textbf{\color{myred} - -}) . 
		In addition, three individual pairs of magnetic field (shim) coils are mounted on corresponding mechanical support structure, marked with (\textbf{\color{myorange}--}). 
		The shim coils enable fine tuning of $\mathbf{B}_0$ along longitudinal and orthogonal (vertical and horizontal) directions. 
		Preparation and detection laser beams ({\color{myblue} $\mathbf{\rightarrow}$}) enter the chamber along $\mathbf{B}_0$. 
		A home-build biquad antenna (sketched in the top left corner) is used to apply microwaves around $2\pi\,\times\,1,600\,$MHz for internal state manipulation of the atom. 
		\textbf{(b)} The magnetic-field variation of the solid-state magnets close to their geometrical centre (inset shows larger region) along $\hat{z}$, calculated using Eq.\,\ref{eq:mag}.
		From numerical calculations, considering all directions, we infer a diameter of spherical volume $d_{\text{dsv}} \simeq 150\,$\textmu m, where $\Delta B_0/B_0 \leq 1\,\times\,10^{-6}$. 
		}
		\label{fig:expsetup}
	\end{figure}

	We equip our experimental setup with a combination of two sets of rare-earth ring magnets and three pairs of field coils (electro magnets) to generate, tune, and stabilise a quantisation field at a strength $\left |\,\mathbf{B}_{0}\,\right |\,\simeq 10.9$\,mT. 
	In Figure\,\ref{fig:expsetup}a, we sketch the geometry of this hybrid setup.
	Each set of the solid-state magnets consists of three neodymium (an alloy made of neodymium, iron, and boron) ring magnets that are axially magnetised.
	Each ring has the following dimensions: $58$\,mm inner diameter, $102$\,mm outer diameter, and $4$\,mm thickness. 
	The vendor specifies the grade of this neodymium in-stock item to be N35, which corresponds to a remanence of $B_{r} \simeq 1.17$\,T and a temperature coefficient of $\simeq -1.2\,\times\,10^{-3}$\,K$^{-1}$\,\cite{die-magnetprofis}.
	We numerically calculate the spatial magnetic field distribution of both sets that are aligned collinear at a distance $d \simeq 223$\,mm (distance between facing planes) using the open-source software package RADIA\,\cite{Elleaume1998,Chubar1998}.
	Along their symmetry axis $\hat{z}$, we can also analytically estimate the field distribution. 
	The magnetic-field strength of a single axially magnetised ring is given by\,\cite{Peng2004}:
	\begin{eqnarray}
	B_{\rm{ring}}(\hat{z}) = \frac{B_r}{2}  \left[
	\left(
	\frac{\hat{z}}{\sqrt{R_{\rm{o}}^2+\hat{z}^2}} 
	-\frac{\hat{z}-D}{\sqrt{R_{\rm{o}}^2+(\hat{z}-D)^2}} 
	\right)
	-\left(
	\frac{\hat{z}}{\sqrt{R_{\rm{i}}^2+\hat{z}^2}} 
	-\frac{\hat{z}-D}{\sqrt{R_{\rm{i}}^2+(\hat{z}-D)^2}} 
	\right) 
	\right],
	\label{eq:mag}
	\end{eqnarray}
	with inner radius, $R_{\rm{i}}$, outer radius, $R_{\rm{o}}$, and thickness $D$.
	We calculate the corresponding field for our magnet assembly by summation of Eq.\,\ref{eq:mag}, geometrically offset for each ring.
	Results of our calculations are shown in Fig.\,\ref{fig:expsetup}b and we, further, numerically estimate the field homogeneity of our magnet configuration in the central region between both sets. 
	Following, we find a diameter of spherical volume $d_{\text{dsv}} \simeq 150\,$\textmu m, where the relative strength of the magnetic field varies less than $1\,\times\,10^{-6}$. 
	In our setup, we mount each set on a threaded cylinder (one turn equals one millimetre travel) to fine tune $d$.
	In this way, we can coarsely tune $|\mathbf{B}_0|$ by $\simeq 0.11$\,mT\,mm$^{-1}$.
	For fine tuning of the spatial alignment and the strength, as well as, temporal stabilisation of $|\mathbf{B}_0|$, we deploy the three pairs of field coils (shim coils). 
	All coil pairs can be fed by current-stabilised low-power supplies with a vendor-specified stability of $0.2\,\times\,10^{-6}$\,A and a maximum current of $0.1$\,A.
	Two pairs can be used for spatial fine tuning and are aligned transversally to $\hat{z}$: the first pair creates a magnetic field of $\simeq 0.24$\,mT for a current of $1$\,A in the horizontal direction and the second pair tunes the vertical direction with $1.3$\,mT A$^{-1}$. 
	The third pair of shim coils is aligned along $\hat{z}$ and we can apply a field strength of $0.26$\,mT A$^{-1}$. 
	In addition, we control the current running in the longitudinal shim coils with our data acquisition system and a resolution of $3\,\times\,10^{-6}$A.
	
	Our experimental apparatus for trapping and controlling single atoms is located in a $\simeq$100 m$^2$ laboratory space that is temperature stabilised to better than $\pm0.3$\,K.
	We trap individual \mgion\ atoms under ultra-high vacuum conditions with a background gas pressure of below
$2\,\times\,10^{-9}$\,Pa in a surface-electrode ion trap. 
	The trap is microfabricated by Sandia National Laboratories and copies of the trap have been previously described\,\cite{Brady2011, Allcock2012}.
	A maximum zero-to-peak voltage $U_\mathrm{RF}\,\simeq\,80$\,V oscillating at $\Omega_\mathrm{RF}\,/(2\pi) \simeq 57.3$\,MHz is applied to two $\simeq\,2.5$\,mm long radio-frequency (RF) electrodes that are $60\,$\textmu m wide and are separated by $\simeq 210 \,$\textmu m.
	This provides confinement of ions in the $x$-$y$ (radial) plane at a distance $h \simeq 83\,$\textmu m above the surface.
	Further, electric (control) potentials are applied to several additional electrodes, in order to confine ions along the $z$ (axial) direction. 
	Correspondingly, we find single-ion motional frequencies of $\simeq\,2\pi\,\times\,0.8$\,MHz (axially) and $\simeq\,2\pi\,\times\,2.1$\,MHz (radially).
	
\begin{figure}
		\centering
		\includegraphics[]{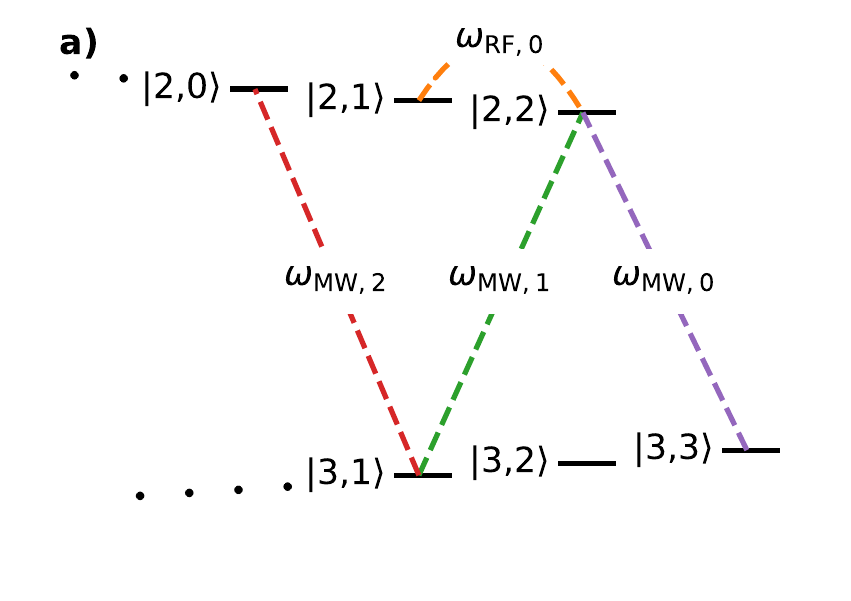}
		\includegraphics[]{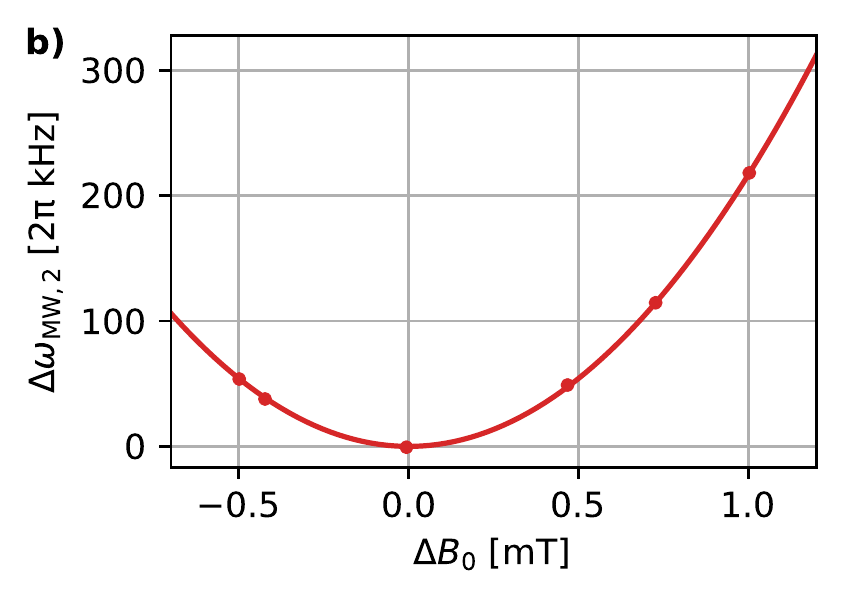}
		\caption{
		\textbf{Ground-state hyperfine level scheme and coarse tuning of the quantisation field.} 
		\textbf{(a)} Relevant Zeeman sub levels of the S$_{1/2}$ hyperfine ground state of $^{25}$Mg$^+$ with a nuclear spin of $5/2$ near $|\mathbf{B}_0|\,\simeq\,10.9$\,mT. 
		In our experiments, we use the marked transitions for internal state manipulation via pulsed microwave or radio-frequency radiation, cp. Tab.\,\ref{tab:coh}.
		\textbf{(b)} Measured (\textbf{\color{myred} \textbullet}) and calculated variation $\Delta\omega_{\rm{MW,\ 2}}$ of the $\left|3,1\right\rangle \leftrightarrow \left|2,0\right\rangle$ transition frequency $\omega_{\rm{MW,\ 2}}/(2\pi) \simeq 1762.974\,$MHz as a function of $\Delta B_0$. 
		Measurement uncertainties are smaller than the markers.
		Tuning of the magnetic field strength (on site of a single trapped atom) via variation of the distance between the solid-state magnets within a full span of $\Delta d \simeq 16\,$mm.
		The local magnetic field strength is probed with a relative precision of better than $\simeq\,0.2\,\times\,10^{-4}$ via the $\left|3,3\right\rangle$ to $\left|2,2\right\rangle$ transition frequency, $\omega_{\rm{MW,\ 0}}$.
		}
		\label{fig:hyperfine}
	\end{figure}

	The external quantisation field is aligned at an angle of approximately $30^{\circ}$ with respect to the $z$ axis and lies within the $x$-$z$ plane (see Fig.\,\ref{fig:expsetup}a).
	In Figure\,\ref{fig:hyperfine}a, we illustrate the level scheme of the $^2$S$_{1/2}$ ground state manifold of \mgion\ with a nuclear spin of $5/2$.
	Near the field strength $|\mathbf{B}_0|\,\simeq\,10.9\,$mT, the $\left|F = 3, m_{F} = 1\right>$ to $\left|F = 2,m_{F} = 0\right>$ hyperfine transition frequency $\omega_{_{\rm{MW,\ 2}}} /(2\pi) \simeq 1,762.974$\,MHz is first-order insensitive to magnetic field changes, while the quadratic frequency deviation is $ \simeq 2\pi\,\times\,217$~kHz mT$^{-2}$. 
	Here, $F$ denotes the total angular momentum and $m_{F}$ is the projection of the angular momentum along the magnetic field axis. 
	We keep this notation for labelling purposes only. In case of $|\mathbf{B}_0| \gg 0$, $F$ and $m_{F}$ are inappropriate quantum numbers and, therefore, we calculate level splittings and inter-state coupling strengths numerically.
	
	Laser beams (with wavelengths close to 280 nm and $\sigma^{+}$-polarised) for Doppler cooling to a temperature of $\simeq 1$\,mK and state preparation via optical pumping into $\left|3,3\right>$ of the $^2$S$_{1/2}$ ground state propagate parallel to the magnetic field. 
	For state detection, a single laser beam induces resonant fluorescence and we can discriminate the $\left|3,3\right>$ (bright) state from the other hyperfine ground (dark) states. 
	Fluorescence photons are detected by a photon-multiplier tube (PMT) detector; more details on our laser setups, state preparation and detection techniques are described in\,\cite{Leibfried2003a, friedenauer2006, Mielenz2016, clos2016, Kalis2016}.
	Further, we can coherently manipulate the internal states via a pulsed application of microwaves between $\omega_{\rm{MW}}/(2\pi) \simeq 1,300\,$MHz and $\simeq 1,850\,$MHz or radio-frequency waves at $\omega_{\rm{RF}}/(2\pi) \simeq 55.3\,$MHz. The microwaves are applied via a home-build biquad antenna\,\cite{Tec2012} that is geometrically optimised for $2\pi\,\times\,1,600\,$MHz, while the radio-frequency waves are capacitively coupled onto the RF electrodes.
	
	Individual experimental sequences are comprised by about $500\,$\textmu s of cooling and state preparation, zero to $1.5\,$s of state manipulations or (near) free evolution, and $100\,$\textmu s of state detection. 
	Sequences are repeated $N_{\rm{exp}} \simeq 100$ to $500$ times to yield averaged data points (including statistical uncertainties) for fixed parameter settings. 
	More details on raw data analysis in our experiments can be found in\,\cite{clos2016}.
	Note, in the following experiments, state preparation can include population transfer from the bright state to any other state of the hyperfine manifold, e.g., $\left|3,1\right>$ state, via microwave (or radio-frequency) pulses. 
	In turn, state detection, will then include reversed application of pulses to transfer population back into the bright state.
	After optimisation, we further neglect infidelities of these transfer pulses in the analysis of our experiments; in similar experimental setups infidelities below $10^{-4}$ have been reported\,\cite{Brown2011a}.
	\section*{Results}	
	\subsection*{Tuning and long-term stability of the quantisation field}
	
	In dedicated calibration measurements, we tune the orientation and strength of $\mathbf{B}_0$, to enable optimal experimental conditions: 
	We require, firstly, first-order field insensitivity of the $\left|3,1\right>$-$\left|2,0\right>$ state splitting and, secondly, optimal state preparation in our experiments. 
	For these calibration experiments, we probe the magnetic field with a single ion via the $\left|3,3\right>$ to $\left|2,2\right>$ transition frequency $\omega_{\rm{MW,\ 0}}$, with a field sensitivity of $\simeq\,-2\pi\,\times\,21.764$\,MHz mT$^{-1}$, cp. Fig.\,\ref{fig:hyperfine} and Tab.\,\ref{tab:coh}. 
	We apply either a single microwave $\pi$ pulse (Rabi sequence, i.e., full population transfer from $\left|3,3\right>$ to $\left|2,2\right>$) or two $\pi/2$ pulses separated by the duration $T_{\rm{Ramsey}} \leq 20\,$\textmu s (Ramsey sequence).

	A coarse setup of the orientation of $\mathbf{B}_0$, i.e., superposition of the magnetic field with the wave vector of our laser beams for optimal optical pumping into the $\left|3,3\right>$ state, is ensured by mechanical/geometrical constrains and adjustments of the beam polarisation.
	Further, we coarsely tune the strength of $\mathbf{B}_0$ by mechanical adjustments of $d$, while monitoring $\omega_{\rm{MW,\ 0}}$ via Rabi sequences.
	In addition, we record $\omega_{\rm{MW,\ 2}}$ via Rabi sequences to find the field strength corresponding to the first-order field-independent transition, see Fig.\,\ref{fig:hyperfine}b. 
	
	For fine tuning of $\mathbf{B}_0$, we adjust current amplitudes fed into the shim coils guided by Ramsey sequences probing $\omega_{\rm{MW,\ 0}}$ in multiple iterations: 
	The currents in the vertical and horizontal shim coils are adjusted to minimise $|\mathbf{B}_0|$, i.e., optimising superposition of $\mathbf{B}_0$ with preparation laser beams, while the current in the longitudinal shim coils is optimised for setting $|\mathbf{B}_0|$ to its target value within a relative precision of $\leq 0.1\,\times\,10^{-4}$.
	We perform multiple long-term measurements of the passive magnetic-field stability over the course of up to 8 hours with a single ion without re-loading or other systematic variations of experimental parameters. We find maximal variations of the magnetic field strength of $\simeq\,0.3\,\times\,10^{-4}$ within five minutes and $\simeq\,1.0\,\times\,10^{-4}$ within one hour.
	During the following measurement runs, we track magnetic-field strength drifts every five to 20 minutes via variations of $\omega_{\rm{MW,\ 0}}$ within a Ramsey sequence, and readjust current amplitudes of the longitudinal shim coils, accordingly.
	In this way, we stabilise the magnetic field to $|\mathbf{B}_0 | = 10.9584(2)$\,mT.
	
	Finally, we conservatively estimate spatial magnetic field gradients in the vicinity of a single trapped ion from final mechanical setup tolerances and based on the numerical field simulations of the solid-state magnets. We assume that the ion is displaced by less than $2\,$mm from the geometric centre position $\hat{z}=0$ of the magnet assembly. Therefore, we expect spatial gradients of less than $11\,$nT \textmu m$^{-1}$ in any direction. Note, this corresponds to a spatial variation of $\omega_{\rm{MW,\ 2}}$ of less than $2\,\pi\,\times\,26\,$\textmu Hz \textmu m$^{-2}$.
	\subsection*{Measurements of coherence times}
\begin{figure}
		\centering
		\includegraphics[]{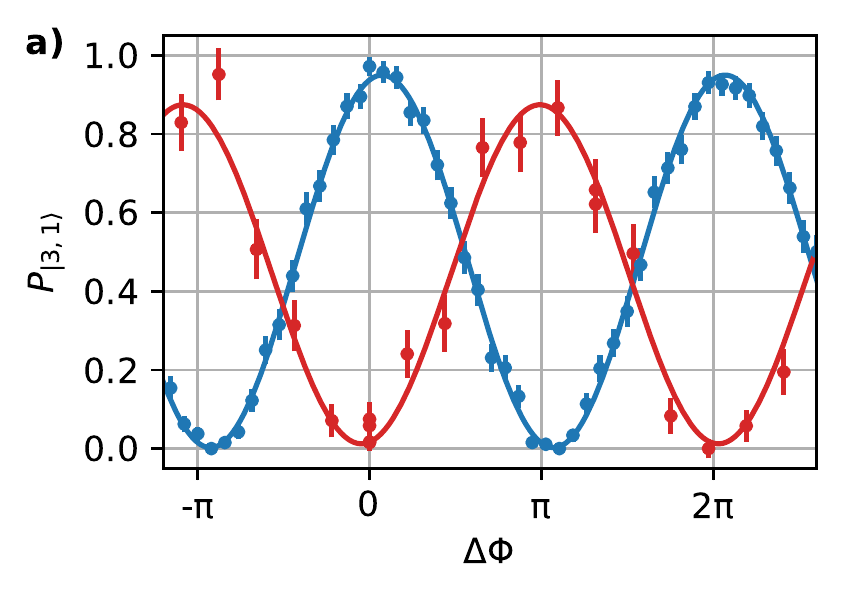}
		\includegraphics[]{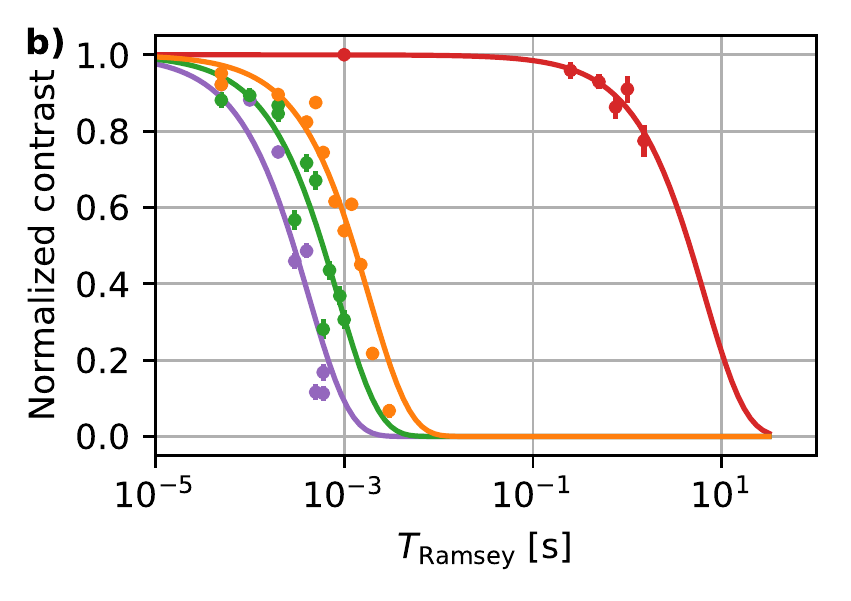}
		\caption{
		\textbf{Measurements of superposition-state coherence times via Ramsey spectroscopy} 
		\textbf{(a)} As an example, we show the variation of state population $P_{\left|3,1\right>}$ as a function of relative (microwave) phase $\Delta\phi$ between two $\pi/2$ pulses, that are separated by $T_{\rm{Ramsey}} = 0.001\,$s (\textbf{\color{myblue}\textbullet}) and $1.0\,$s (\textbf{\color{myred}\textbullet}); error bars correspond to statistical uncertainties (s.e.m.).
		Sinusoidal model fits to both data give best values for the achieved contrast of $0.948(7)$ and $0.86(4)$, respectively. 
		\textbf{(b)} Evolution of contrast normalized to the initial contrast for four different stets of internal state superpositions using states 
		$\left|3,1\right>$-$\left|2,0\right>$ (\textbf{\color{myred}\textbullet}), 
		$\left|2,1\right>$-$\left|2,2\right>$ (\textbf{\color{myorange}\textbullet}), 
		$\left|3,1\right>$-$\left|2,2\right>$ (\textbf{\color{mygreen}\textbullet}), 
		and $\left|3,3\right>$-$\left|2,2\right>$ (\textbf{\color{myviolet}\textbullet}), 
		as a function of $T_\mathrm{Ramsey}$; error bars indicate s.e.m..
		Individual fits of exponential decays to each data set yield coherence times $\tau$, defined as the duration where contrast reaches a level of $e^{-1}$. 
		We find corresponding durations of 
		$6.6(0.9)\,$s (\textbf{\color{myred}--}), 
		$0.0018(2)\,$s (\textbf{\color{myorange}--}), 
		$0.0009(1)\,$s (\textbf{\color{mygreen}--}), 
		and $0.00042(6)\,$s (\textbf{\color{myviolet}--})
		for the four different superposition states, respectively.
		}
		\label{fig:ramsey}
	\end{figure}

	In the following, we determine coherence times $\tau$ of four different stets of internal state superpositions within the ground state hyperfine manifold, cp. Fig.\,\ref{fig:hyperfine}a, in order to further benchmark the performance of our overall setup. 
	In Table\,\ref{tab:coh}, we quantify and summarise relevant properties of the probed transitions.
	\begin{table}
	\begin{center}
	\begin{tabular}{cccccc}
	\hline
	\hline
	 & Trans. frequency & 	Field sensitivity& 	Coupling strength &	Coherence time \\
	 & 	[$2\pi$\,MHz]& [MHz\,mT$^{-1}$]& [$2\pi$\,kHz]	&	$\tau$ [s]\\
	\hline
	 MW, 0& $1541.066(4)$& $-21.764$& $161(3)$&	$0.42(6)\,\times\,10^{-3}$\\
 	 MW, 1& $1655.815(2)$& $-10.116$& $38.3(8)$&	$0.9(1)\,\times\,10^{-3}$\\ 	 
	 MW, 2& $1762.97381160(1)$& $\pm 0(1)\,\times\,10^{-4}$ ($+0.217\,$mT$^{-1}$)& $28.5(6)$& $6.6(9)$ \\
	 RF, 0& $55.260(1)$& $+5.381$& $0.286(6)$ & $1.8(2)\,\times\,10^{-3}$\\	
	\hline
	\hline	
	\end{tabular}
	\end{center}
	\caption{
	\textbf{Properties of the four probed hyperfine transitions.} 
	Calculated transition frequencies and magnetic field sensitivities are listed, as well as, a summary of the experimentally applied coupling strengths and measured coherence times. All values are taken for a measured magnetic field of $|\mathbf{B}_0 |= 10.9584(2)$\,mT
	}
	\label{tab:coh}
	\end{table}%
	We apply the following experimental sequences to measure coherence times: 
	After preparation of the initial state, we create internal state superposition states via a first $\pi/2$ (microwave or radio-frequency) pulse, wait for fixed durations $T_{\rm{Ramsey}}$, apply a second $\pi/2$ pulse with variable phase $\Delta\phi$ (relative to the phase of the first pulse) and detect the final state. 
	In Figure~\ref{fig:ramsey}a, we show, as an example, results of the field-independent superposition states.
	We plot the population probability $P_{\left|3,1\right>}$ of state $\left|3,1\right>$ as a function of $\Delta\phi$ for two different values of $T_{\rm{Ramsey}}$. 
	From sinusoidal model fits to the data, we determine the contrast of such Ramsey sequences for all four sets of superposition states for variable $T_{\rm{Ramsey}}$ and show these results in Fig.\,\ref{fig:ramsey}b.
	In a final analysis step, we determine $\tau$, i.e., the duration $T_{\rm{Ramsey}}$ after which the initial contrast decayed to $e^{-1}$, by exponential model fits to each data set.
	We find $\tau_{\rm{MW,\ 2}} = 6.6(9)\,$s for the field-independent superposition states, while coherence times are shorter than two milliseconds for all other superposition states; all results are summarised in Tab.\,\ref{tab:coh}.
	Note, we ensure that leakage from our preparation laser beams contribute less than $2\pi\,\times\,0.08$\,Hz (for all probed transitions).
	Measured decoherence rates $\Gamma = 2\pi\,\tau^{-1}$ increase linearly as a function of the corresponding magnetic-field sensitivities and suggesting significant magnetic-field fluctuations on time scales between a few hundred microseconds and a few seconds.
	From additional experiments with less stable power supplies feeding the shim coils, we estimate that noise levels from the relevant current supplies contribute less than $2\pi\,\times\,0.002$\,Hz to the lowest decoherence rates of $\Gamma_{\rm{MW,\ 2}} / (2\pi) = 0.15(3)\,$Hz. 
	In turn, we assume fluctuations from stray magnetic fields to dominate magnetic noise fields.
	\subsection*{Sensing of oscillating magnetic fields}
\begin{figure}
\centering
\includegraphics[]{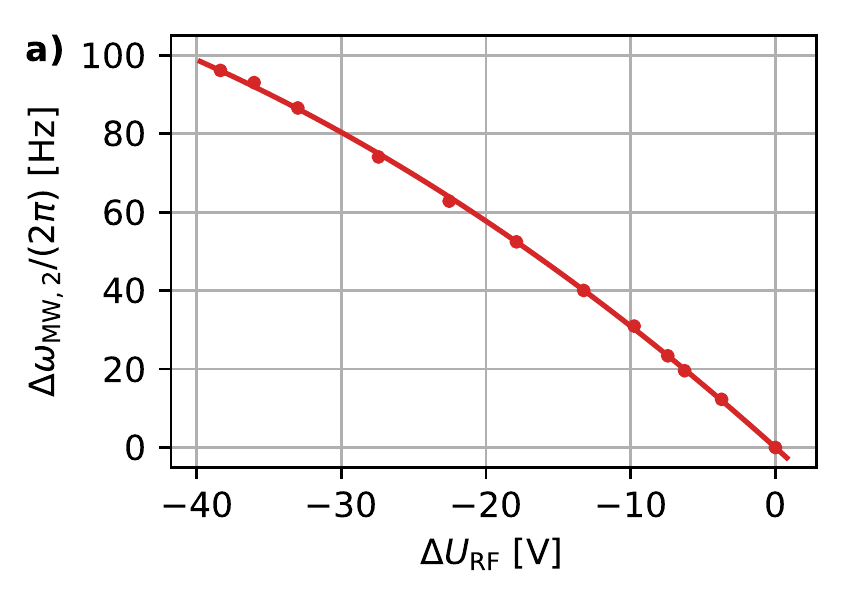}
\includegraphics[]{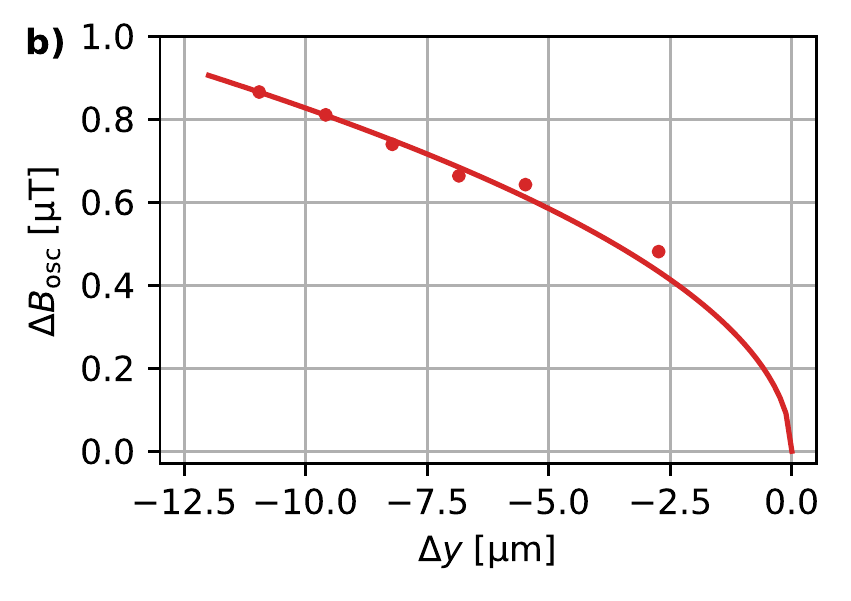}
\caption{
\textbf{Sensing of oscillating magnetic fields  originating from stray radio-frequency currents.}
Detection of variations of field strength $B_{\rm{osc}}$ via spin-echo sequences that record phase accumulation from induced differential (frequency dependent) a.c. Zeeman shifts, while the ion is in a superposition state of $\left|3,1\right\rangle$ and $\left|2,0\right\rangle$. 
For fields oscillating at $\Omega_{\rm{osc}} = \Omega_{\rm{RF}} = 2\pi\,\times\,57.3\,$MHz and our setup, we calculate a quadratic sensitivity of $2\pi\times4.783\,$Hz \textmu T$^{-2}$ .
\textbf{(a)} Differential a.c. Zeeman shifts $\Delta\omega_{\rm{MW,\ 2}}$ as a function of $\Delta U_\mathrm{RF}$, i.e., systematic variation of stray currents $\propto U_\mathrm{RF}$. 
Uncertainties of data points are smaller than the markers.
We find a slope of $2\pi\times20.77(7)\,\rm{mHz\,V}^{-2}$ from a model fit and conclude $B_{\rm{osc}} = 5.239(8)\,$\textmu T for our typical operation condition $U_{\rm{RF}} = 79.5\,$V. 
\textbf{(b)} Spatial variation of $B_{\rm{osc}}$ along the $y$ direction. 
Data is in agreement with a model fit with a slope of $\delta B_{\rm{osc}}/\sqrt{\delta y} = 0.262(3)\,$\textmu T \textmu m$^{-1/2}$.
}
\label{fig:acz}
\end{figure}

	In a first application, we use the clock transition for sensing of oscillating magnetic fields $\mathbf{B}_{\rm{osc}}$ that originate from stray currents with unknown amplitude ($\propto\,U_{\rm{RF}}$) in the two radio-frequency electrodes. 
	We consider that these fields predominantly lie in the $x$-$y$ plane, due to the symmetry of the electrode structure.
	Under this assumption and from basic atomic properties, we calculate the frequency dependent a.c. Zeeman shift\,\cite{warring2013a} of the probed transition, and find a quadratic sensitivity of $2\pi\times4.783\,$Hz \textmu T$^{-2}$ to fields oscillating at $\Omega_{\rm{RF}}$. 
	To further characterise $\mathbf{B}_{\rm{osc}}$, we apply the following experimental (spin-echo) sequence and detect phase accumulations from differential a.c. Zeeman shifts due to a variation of $U_{\rm{RF}}$:
	After preparation of $\left|3,1\right>$, a $\pi/2$ pulse to create a $\left|3,1\right>$-$\left|2,0\right>$ superposition, and a free evolution duration $T_{P}$, we apply a $\pi$ pulse in phase with the previous pulse. 
	After an additional duration $T_{\rm{P}}$, during which we ramp down (and back) the radio-frequency voltage by $\Delta U_{\rm{RF}}$ within ramp durations of $\leq 80\,$\textmu s, we conclude the experimental sequence with a second $\pi/2$ pulse (again, in phase with the previous pulses) and detection of the $\left|3,1\right>$ state.
	Note, the spin echo sequence makes results insensitive to quantisation-field fluctuations slower than the time scale of an individual sequence ($\simeq 1$\,s).
	In subsequent measurements, we vary $T_{\rm{P}}$ to up to $1.2$\,s for fixed $\Delta U_{\rm{RF}}$ to determine the phase accumulation from differential a.c. Zeeman shifts. 
	Corresponding results as a function of $\Delta U_{\rm{RF}}$ are shown in Fig.~\ref{fig:acz}a.
	A quadratic model fit to this data yields a slope of $2\pi\times20.77(7)\,\rm{mHz\,V}^{-2}$ and from this we infer an oscillating magnetic field strength of $B_{\rm{osc}} = 5.239(8)\,$\textmu T for $U_{\rm{RF}} = 79.5\,$V. 
	
	Next, we measure the spatial dependence of this field along the $y$ axis. 
	We deploy a similar spin-echo sequence as described above, but vary the position $\Delta y$ of the ion within the second free evolution duration for fixed $U_{\rm{RF}}$.
	In such sequences, the ion position is varied by applying electric control fields. 
	We calibrate relative ion displacements in dedicated measurements to within $\pm 0.2\,$\textmu m and ensure that displacements in all other directions are less than $\simeq\,1.0\,$\textmu m for maximal $y$ displacements.
	We observe a linear variation of $\Delta \omega_{\rm{MW,\ 2}}$ as a function of $\Delta y$ with a slope of $2\pi\times327(7)$\,mHz\,\textmu m$^{-1}$ and attribute this to differential a.c. Zeeman shifts.
	Note, to explain the observed frequency shift by a spatial variation of static magnetic fields only, it would require local gradients of $\simeq1.2\,$mT \textmu m$^{-1/2}$. 
	In comparison, we refer to our estimation of global linear gradients of less than $11\,$nT \textmu m$^{-1}$ (see above) and judge the presence of such large (static) local gradients to be unlikely in our setup.
	Consequently, we show in Figure\,\ref{fig:acz}b the variation of $B_{\rm{osc}}$ with a non-linear slope of $\delta B_{\rm{osc}}/\sqrt{\delta y} = 0.261(3)\,$\textmu T \textmu m$^{-1/2}$.

	\section*{Discussion}
	We describe a hybrid approach for generating stable magnetic fields with a field strength around $10.9$\,mT and a spatial variation of less than 10$^{-6}$ within a diameter of spherical volume of $150\,$\textmu m, using a combination of rare-earth magnets and magnetic field coils powered by stable low-power current supplies. 
	We coarsely tune the magnetic field by mechanical adjustments of the permanent magnets and use the field coils for fine tuning. 
	In our experiments, we use a single trapped Mg$^+$ atom to probe the field characteristics.
We find a passive long-term temporal stability of $\simeq 1\,\times\,10^{-4}$ over the course of one hour. 
	In addition, we implement a feed-back loop for active field stabilisation to better than $2\,\times\,10^{-5}$ via re-adjustments of currents in the field coils every five to 20 minutes. 
	Further, we benchmark the short-term performance of our setup by measurements of coherences of
internal state superpositions and find coherence times of up to $6.6(9)$s. 
	We infer that short-term stability is limited by fluctuations of stray magnetic fields. 
	In a first quantum sensing application, relying on such high performance, we probe magnetic fields oscillating at $2\pi\,\times\,60\,$MHz that originate from currents running in our trapping structure. 
	We measure the magnitude with a quadratic sensitivity of $2\pi\,\times\,4.783\,$Hz mT$^{-2}$ and spatial variation within about ten micrometers. 
	In an extension of our measurements, complete, i.e., local amplitude and phase information of the oscillating field can be recorded\,\cite{Bohi2010}.
	Numerical simulations of the oscillating magnetic fields can be compared to our results and, in turn, would yield detailed understanding of electronic properties of trapping structures that are used for quantum simulation\,\cite{Mielenz2016} and related fields of research.
	
	In the future, the passive stability can be further improved via implementation of shielding against stray magnetic fields as, e.g., demonstrated in \cite{Ruster2016c}. 
	Further, depending on the environment conditions, the use of Samarium Cobalt magnets that are less sensitive to temperature variations can be recommended \cite{Ruster2016c}.
	Adapted and optimised geometries of the solid-state magnets can yield smaller footprints, while increasing regions of homogenous field distribution, and variable field strengths. 
	This is of particular importance for applications that have strict power and load requirements and can be more cost effective.
	We conclude that our approach can contribute to developments of more compact and robust experimental setups in a variety of applications. 
		
	\section*{Data availability}
	
	The datasets generated and analysed during the current study are available from the corresponding author on reasonable request.
	
	\bibliography{refs}
	
	\section*{Acknowledgements}
	
	We thank Frederik Jacobs for introducing us to the RADIA software package. The article processing charge was funded by the German Research Foundation (DFG) and the University of Freiburg in the funding programme Open Access Publishing.
	
	\section*{Author contributions statement}
	
	F.H., P.K., M.W., U.W., and T.S. developed the experimental apparatus. F.H., P.K., and U.W. conceived, conducted, and analysed the experiments. All authors carefully discussed the results and reviewed the manuscript.

	\section*{Competing financial interests}
	
	The authors declare that they have no competing financial interests.

\end{document}